\documentclass[aps,prl,twocolumn,groupedaddress,showpacs]{revtex4-1}

\usepackage{graphicx}% Include figure files

\begin{document}

\parbox{\textwidth}{\center (Physical Review Letters - {\it{in press}})}
\\
\newline

\title{The Beam-Target Helicity Asymmetry for $\vec{\gamma} \vec{n} \rightarrow \pi^-  p$ in the {\bf{$N^*$}} Resonance Region}

%%%%%%%%%%%%%%%  institute addresses  %%%%%%%%%%%%%%%%%%%%%%%%% 
\newcounter{univ_counter}
\setcounter{univ_counter} {0}

\addtocounter{univ_counter} {1} 
\edef\JLAB{$^{\arabic{univ_counter}}$ }

\addtocounter{univ_counter} {1} 
\edef\CMU{$^{\arabic{univ_counter}}$ }

\addtocounter{univ_counter} {1} 
\edef\CUA{$^{\arabic{univ_counter}}$ }

\addtocounter{univ_counter} {1} 
\edef\EDINBURGH{$^{\arabic{univ_counter}}$ }

\addtocounter{univ_counter} {1} 
\edef\GWU{$^{\arabic{univ_counter}}$ }

\addtocounter{univ_counter} {1} 
\edef\JMU{$^{\arabic{univ_counter}}$ }

\addtocounter{univ_counter} {1} 
\edef\BONN{$^{\arabic{univ_counter}}$ }

\addtocounter{univ_counter} {1} 
\edef\NSU{$^{\arabic{univ_counter}}$ }

\addtocounter{univ_counter} {1} 
\edef\ODU{$^{\arabic{univ_counter}}$ }

\addtocounter{univ_counter} {1} 
\edef\GATCHINA{$^{\arabic{univ_counter}}$ }

\addtocounter{univ_counter} {1} 
\edef\ZAGREB{$^{\arabic{univ_counter}}$ }

\addtocounter{univ_counter} {1} 
\edef\CLERMONT{$^{\arabic{univ_counter}}$ }

\addtocounter{univ_counter} {1} 
\edef\UCONN{$^{\arabic{univ_counter}}$ }

\addtocounter{univ_counter} {1} 
\edef\IOWA{$^{\arabic{univ_counter}}$ }

\addtocounter{univ_counter} {1} 
\edef\ROMAII{$^{\arabic{univ_counter}}$ }

\addtocounter{univ_counter} {1} 
\edef\VIRGINIA{$^{\arabic{univ_counter}}$ }
%%%
\addtocounter{univ_counter} {1} 
\edef\ANL{$^{\arabic{univ_counter}}$ }

\addtocounter{univ_counter} {1} 
\edef\ASU{$^{\arabic{univ_counter}}$ }

\addtocounter{univ_counter} {1} 
\edef\CSUDH{$^{\arabic{univ_counter}}$ }

\addtocounter{univ_counter} {1} 
\edef\CANISIUS{$^{\arabic{univ_counter}}$ }

\addtocounter{univ_counter} {1} 
\edef\CNU{$^{\arabic{univ_counter}}$ }

\addtocounter{univ_counter} {1} 
\edef\WM{$^{\arabic{univ_counter}}$ }

\addtocounter{univ_counter} {1} 
\edef\FAIR{$^{\arabic{univ_counter}}$ }

\addtocounter{univ_counter} {1} 
\edef\FIU{$^{\arabic{univ_counter}}$ }

\addtocounter{univ_counter} {1} 
\edef\FSU{$^{\arabic{univ_counter}}$ }

\addtocounter{univ_counter} {1} 
\edef\INFNFE{$^{\arabic{univ_counter}}$ }

\addtocounter{univ_counter} {1} 
\edef\INFNFR{$^{\arabic{univ_counter}}$ }

\addtocounter{univ_counter} {1} 
\edef\INFNGE{$^{\arabic{univ_counter}}$ }

\addtocounter{univ_counter} {1} 
\edef\INFNTUR{$^{\arabic{univ_counter}}$ }

\addtocounter{univ_counter} {1} 
\edef\ISU{$^{\arabic{univ_counter}}$ }

\addtocounter{univ_counter} {1} 
\edef\ORSAY{$^{\arabic{univ_counter}}$ }

\addtocounter{univ_counter} {1} 
\edef\ITEP{$^{\arabic{univ_counter}}$ }

\addtocounter{univ_counter} {1} 
\edef\SACLAY{$^{\arabic{univ_counter}}$ }

\addtocounter{univ_counter} {1} 
\edef\KNU{$^{\arabic{univ_counter}}$ }

\addtocounter{univ_counter} {1} 
\edef\MISS{$^{\arabic{univ_counter}}$ }

\addtocounter{univ_counter} {1} 
\edef\OHIOU{$^{\arabic{univ_counter}}$ }

\addtocounter{univ_counter} {1} 
\edef\TEMPLE{$^{\arabic{univ_counter}}$ }

\addtocounter{univ_counter} {1} 
\edef\MSU{$^{\arabic{univ_counter}}$ }

\addtocounter{univ_counter} {1} 
\edef\GLASGOW{$^{\arabic{univ_counter}}$ }

\addtocounter{univ_counter} {1} 
\edef\UNH{$^{\arabic{univ_counter}}$ }

\addtocounter{univ_counter} {1} 
\edef\URICH{$^{\arabic{univ_counter}}$ }

\addtocounter{univ_counter} {1} 
\edef\SCAROLINA{$^{\arabic{univ_counter}}$ }

\addtocounter{univ_counter} {1} 
\edef\UTFSM{$^{\arabic{univ_counter}}$ }

\addtocounter{univ_counter} {1} 
\edef\VT{$^{\arabic{univ_counter}}$ }

\addtocounter{univ_counter} {1} 
\edef\YEREVAN{$^{\arabic{univ_counter}}$ }

%%%%%%%%%%%%%%%%%%%%%

\author{
D. Ho,\CMU\
P. Peng,\VIRGINIA\
C. Bass,\JLAB\
P. Collins,\CUA\
A. D'Angelo,\JLAB$\!\!^,$\ROMAII\ 
A. Deur,\JLAB\
J. Fleming,\EDINBURGH\
C. Hanretty,\JLAB$\!\!^,$\VIRGINIA\ \\
T. Kageya,\JLAB\
M. Khandaker,\NSU\
F.J. Klein,\GWU
% ...
\\
(The CLAS Collaboration)
}
%%%%%%%%%%%%%%%%%%%%%

\affiliation{\JLAB Thomas Jefferson National Accelerator Facility, Newport News, Virginia 23606, USA}
\affiliation{\CMU Carnegie Mellon University, Pittsburgh, Pennsylvania 15213, USA}
\affiliation{\CUA Catholic University of America, Washington, D.C. 20064, USA}
\affiliation{\EDINBURGH Edinburgh University, Edinburgh EH9 3FD, United Kingdom}
\affiliation{\GWU The George Washington University, Washington, DC 20052, USA}
\affiliation{\JMU James Madison University, Harrisonburg, Virginia 22807, USA}
\affiliation{\BONN Helmholtz-Institut f\"ur Strahlen- und Kernphysik, Universit\"at Bonn, 53113 Bonn, Germany}
\affiliation{\NSU Norfolk State University, Norfolk, Virginia 23504, USA}
\affiliation{\ODU Old Dominion University, Norfolk, Virginia 23529, USA}
\affiliation{\GATCHINA Petersburg Nuclear Physics Institute, Gatchina, 188300 Russia}
\affiliation{\ZAGREB Rudjer Bo\v{s}kovi\'{c} Institute, Zagreb 10002, Croatia}
\affiliation{\CLERMONT Universit\'e Blaise Pascal, Clermont-Ferrand, 63178 Aubi\`ere Cedex, France}
\affiliation{\UCONN University of Connecticut, Storrs, Connecticut 06269, USA}
\affiliation{\IOWA University of Iowa, Iowa City, Iowa 52242, USA}
\affiliation{\ROMAII Universit\`a di Roma ``Tor Vergata" and INFN Sezione di Roma2, 00133 Roma, Italy}
\affiliation{\VIRGINIA University of Virginia, Charlottesville, Virginia 22903, USA}
%%%
\affiliation{\ANL Argonne National Laboratory, Argonne, Illinois 60439, USA}
\affiliation{\ASU Arizona State University, Tempe, Arizona 85287, USA}
\affiliation{\CSUDH California State University, Dominguez Hills, Carson, CA 90747, USA}
\affiliation{\CANISIUS Canisius College, Buffalo, New York 14208, USA}
\affiliation{\CNU Christopher Newport University, Newport News, Virginia 23606, USA}
\affiliation{\WM College of William and Mary, Williamsburg, Virginia 23187, USA}
\affiliation{\FAIR Fairfield University, Fairfield, Connecticut 06824, USA}
\affiliation{\FIU Florida International University, Miami, Florida 33199, USA}
\affiliation{\FSU Florida State University, Tallahassee, Florida 32306, USA}
\affiliation{\INFNFE INFN Sezione di Ferrara and Universita' di Ferrara, 44121, Ferrara, Italy}
\affiliation{\INFNFR INFN, Laboratori Nazionali di Frascati, 00044 Frascati, Italy}
\affiliation{\INFNGE INFN, Sezione di Genova, 16146 Genova, Italy}
\affiliation{\INFNTUR INFN, Sezione di Torino, 10125 Torino, Italy}
\affiliation{\ISU Idaho State University, Pocatello, Idaho 83209}
\affiliation{\ORSAY Institut de Physique Nucl\'eaire, CNRS/IN2P3 and Universit\'e Paris Sud, Orsay, France}
\affiliation{\ITEP Institute of Theoretical and Experimental Physics, Moscow, 117259, Russia}
\affiliation{\SACLAY Irfu/SPhN, CEA, Universit\'e Paris-Saclay, 91191 Gif-sur-Yvette, France}
\affiliation{\KNU Kyungpook National University, Daegu 41566, Republic of Korea}
\affiliation{\MISS Mississippi State University, Mississippi State, Mississippi 39762, USA}
\affiliation{\OHIOU Ohio University, Athens, Ohio  45701, USA}
\affiliation{\TEMPLE Temple University,  Philadelphia, Pennsylvania 19122, USA}
\affiliation{\MSU Skobeltsyn Institute of Nuclear Physics, Lomonosov Moscow State University, 119234 Moscow, Russia}
\affiliation{\GLASGOW University of Glasgow, Glasgow G12 8QQ, United Kingdom}
\affiliation{\UNH University of New Hampshire, Durham, New Hampshire 03824, USA}
\affiliation{\URICH University of Richmond, Richmond, Virginia 23173, USA}
\affiliation{\SCAROLINA University of South Carolina, Columbia, South Carolina 29208, USA}
\affiliation{\UTFSM Universidad T\'{e}cnica Federico Santa Mar\'{i}a, Casilla 110-V Valpara\'{i}so, Chile}
\affiliation{\VT Virginia Polytechnic Institute and State University, Blacksburg, Virginia 24061, USA}
\affiliation{\YEREVAN Yerevan Physics Institute, 375036 Yerevan, Armenia}
%%%%%%%%%%%%%%%%%%%%%%%%%%%%%%%%%%%%%%%%%%%%%%%%%%%%%%%%%%%%

\date{\today}

\begin{abstract}
We report the first beam-target double-polarization asymmetries in the $\gamma + n(p) \rightarrow \pi^- + p(p)$ reaction spanning the nucleon resonance region from invariant mass $W$= $1500$ to $2300$ MeV. Circularly polarized photons and longitudinally polarized deuterons in $H\!D$ have been used with the CLAS detector at Jefferson Lab. The exclusive final state has been extracted using three very different analyses that show excellent agreement, and these have been used to deduce the {\it{E}} polarization observable for an effective neutron target. These results have been incorporated into new partial wave analyses, and have led to significant revisions for several $\gamma nN^*$ resonance photo-couplings.

\end{abstract}

%\keywords{}

\maketitle %must follow title, authors, abstract, \pacs, and \keywords

A successful description of the excited levels of a composite system is a basic test of how well the underlying forces are understood. While Quantum Chromodynamics (QCD) is generally regarded as a mature theory of interacting quarks that has been very successful in the asymptotically free regime, the excited states of the nucleon pose many challenges. This partly arises because of the complexity of multiple effects that {\it{dress}} the interactions (such as meson loops and channel couplings \cite{EBAC-P11}, which are beyond the scope of present Lattice QCD \cite{LQCD11}), and partly because the states are broad and overlapping, making their production amplitudes difficult to disentangle without constraints from many different types of measurements \cite{SHKL11}. Until relatively recently, excited baryon resonances had been identified almost exclusively from $\pi N$ scattering data, which yielded only a fraction of the number of levels expected \cite{CQM98, LQCD11}. However, new candidate states have now emerged from analyses of a large number of meson photoproduction experiments \cite{PDG16}. The associated $\gamma NN^*$ electromagnetic couplings in the full spectrum provide a measure of dynamical properties and serve as benchmarks for models of nucleon structure. 

To isolate an excited nucleon state requires a decomposition of the reaction amplitude into multipoles of definite spin, parity, and isospin. Single pseudo-scalar meson photoproduction is described by 4 complex amplitudes and requires data on a minimum of 8 (out of 16) different spin observables to avoid mathematical ambiguities, although in practice even larger numbers are needed to overcome the limitations imposed by experimental accuracy \cite{SHKL11,Sam-NSTAR11}. In recent years, major experimental campaigns have been mounted at several laboratories to measure many different spin asymmetry combinations with proton targets. However, the electromagnetic interaction does not conserve isospin. In particular, the amplitude for the $N(\gamma,\pi)$ reaction factors into distinct isospin components, {\small $\mathcal{A}_{(\gamma,\pi^{\pm})} =\surd{2} \{\mathcal{A}^{I=1/2}_{p/n} \mp {\tiny{1/3~}}\mathcal{A}^{I=3/2}\} $}. Thus, while the excitation of {\small {\it {I}}=3/2} $\Delta^*$ states can be entirely determined from proton target data, measurements with both neutron and proton targets are required to deduce the isospin {\small {\it {I}}=$1/2$} amplitudes, and separate $\gamma pN^*$ and $\gamma nN^*$ couplings. Generally, the latter are poorly determined, due to the paucity of neutron reaction data.

The E06-101 experiment at Jefferson Lab, the {\it{g14}} run with the CEBAF Large Acceptance Spectrometer (CLAS) in Hall B  \cite{CLAS}, has focused on constraining photoproduction amplitudes with new spin observables from polarized neutrons. Here we report the first beam-target double polarization measurements of $E=\frac{1}{P_\gamma P_T} \frac{\sigma_A - \sigma_P}{\sigma_A + \sigma_P}$ in the quasi-free reaction $\gamma~n(p) \rightarrow \pi^- p(p)$, through the $N^*$ resonance region. These have been measured with beam ($P_\gamma$) and target polarizations ($P_T$) anti-parallel ($A$) and parallel ($P$) to the beam momentum, using the sign convention of Ref. \cite{Rosetta12}. 

\begin{figure}
\center{
\includegraphics[scale=0.6]{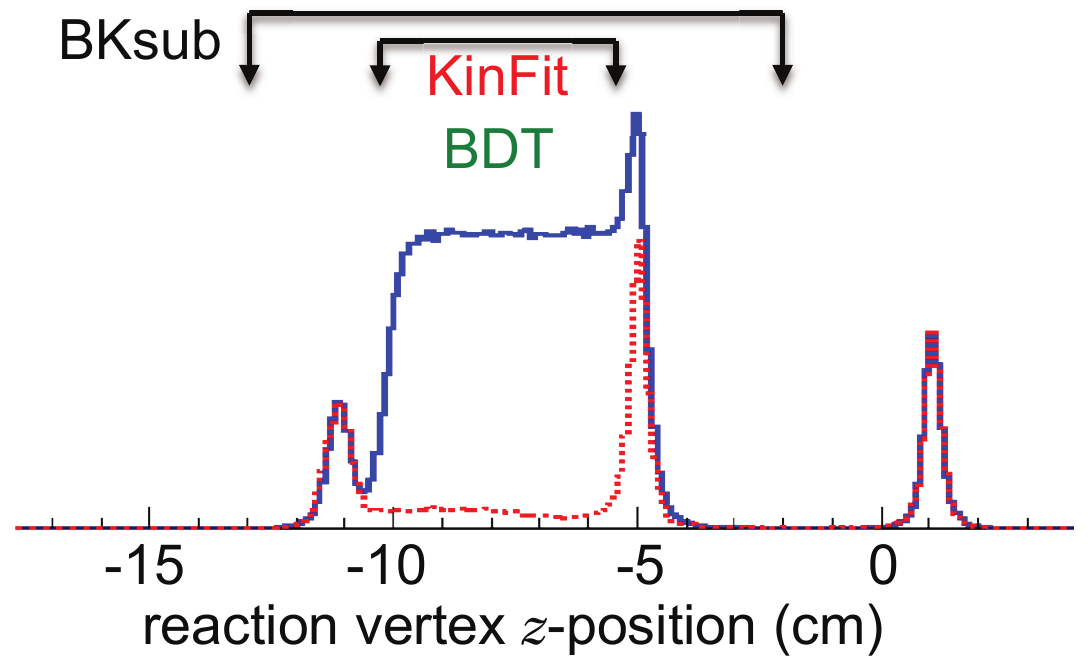}
\caption{Reaction vertex position along the beam direction ($z$), reconstructed by tracking $\pi^-$ and $p$ in CLAS, shown for equivalent-flux data from full ({\it{solid blue}}) and empty ({\it{dotted red}}) targets. Beam entrance and exit windows generate peaks at -11 and -5 cm, respectively. A target-independent foil in the cryostat generates the peak at +1 cm. Regions included in the BKsub, and in the KinFit and BDT analyses are indicated. }
\label{TGTvertexC} }   
\end{figure}

Tagged photons, with circular polarization up to 85\%, were generated by the bremsstrahlung of longitudinally polarized electrons \cite{tagger}, and spanned the energy range from 0.7 to 2.4 GeV. The electron polarization was periodically monitored by M{\o}ller scattering and the helicity transferred to the photon was calculated from Ref. \cite{Max}. The beam polarization was flipped in a semi-random pattern at 960 Hz by flipping the electron helicity, with a charge-flux asymmetry between the two states of less than $10^{-3}$. Photons were incident on 5 cm long frozen-spin targets of longitudinally polarized hydrogen-deuteride ($H\!D$) in the solid phase \cite{HDice1, HDice2, XWice3}. $D$ polarizations were monitored frequently in-beam with NMR \cite{HDice2} and averaged 25\%, with relaxation times in excess of a year. A sample reconstruction of the $\pi^-p$ reaction vertex is shown in Fig.~\ref{TGTvertexC} as the solid (blue) histogram. Background reactions from the unpolarizable material of the target cell, pCTFE[$C_2ClF_3$] walls and Al cooling wires \cite{HDice1}, were small.  These could be directly measured by warming the cell and pumping out the $H\!D$ gas (dotted red histogram). After subtraction, the deuterium of the $H\!D$ provided the only source of neutrons.

In the analysis of E06-101, advanced techniques such as {\it Kinematic Fitting} and {\it Boosted Decision Trees}, have been employed to study other channels with multi-particle final states and/or low cross sections. To validate the implementation of these complex methods, each has been applied to this same high-statistics channel having only charged particles in the final state, $\gamma D\rightarrow\  \pi^- p (p)$. These have been compared to a conventional analysis of sequential one-dimensional selection requirements with empty-target subtraction. This comparison has provided an opportunity to assess possible differences between analysis philosophies. Each analysis selected events with exactly one $\pi^-$ and one $p$, both identified by the correlations between their velocities and momenta in CLAS. 

 In the conventional {\it{Background-Subtraction}} (BKsub) analysis, a sequence of cuts was applied to isolate the final state. Since in the quasi-free limit the desired reaction from the neutron is 2-body, only events with an azimuthal angle difference between the $p$ and the $\pi^-$ of $180^{\circ} \pm 20^{\circ}$ were accepted. The undetected {\it{spectator}} proton of the reaction $\gamma+D\rightarrow \pi^- +p+(p_s)$ was reconstructed and the square of its missing mass was required to be less than 1.1 GeV$^2$. Backgrounds from the target cell, including the beam-entrance and exit windows (as indicated in Fig.~\ref{TGTvertexC}), were subtracted for each kinematic bin using flux-normalized empty-cell data.

{\it{Kinematic fitting}} (KinFit) used the constraints of energy and momentum conservation to improve the accuracy of measured quantities, and so obtained improved estimates on the momenta of undetected particles~\cite{KinFit}. This allowed a separation of reactions with additional particles in the final state, as well as reactions on bound nucleons in the target cell material, since these deviated from elementary kinematics. In this analysis, a pre-selection based on vertex reconstruction was used to eliminated the target cell windows (as in Fig.~\ref{TGTvertexC}). For each event, a confidence level, calculated for the reaction $\gamma + n \rightarrow \pi^- + p$ where the target was assumed to have the neutron mass but unknown momentum \cite{Peng}, was required to be $\geq$ 0.05. This procedure significantly suppressed events from high-momentum neutrons in the deuteron. 

When processing exclusive events, many kinematic variables can be constructed. Conventional BKsub-style analyses view each variable in different projections to one or two dimensions where sequential requirements are placed on data. In contrast, multivariate {\it{Boosted Decision Trees}} (BDT) can be used to view each event in a higher dimension where all requirements can be placed simultaneously \cite{BDT1, BDT2}. The process creates a {\it{forest}} of logical {\it{if-then-else}} tests for all kinematic variables and the resulting decision trees are applied to all of the information. In this application, $\pi^-$+{\it{p}} candidate events are pre-selected and their reconstructed origin is required to lie within a region excluding the target cell windows (Fig.~\ref{TGTvertexC}). The BDT algorithm is {\it{trained}} to select signal events on the results of a Monte Carlo simulation of the CLAS response to the reaction of interest and {\it{trained}} to reject background on the empty-cell data. The algorithm then is used to categorize each reaction event as either {\it{signal}} or {\it{background}} \cite{Dao}. Overall, this procedure retains about 25\% more events (compared with the BKsub analysis), which results in smaller statistical uncertainties.

\begin{figure}
\center{
\includegraphics[scale=0.5]{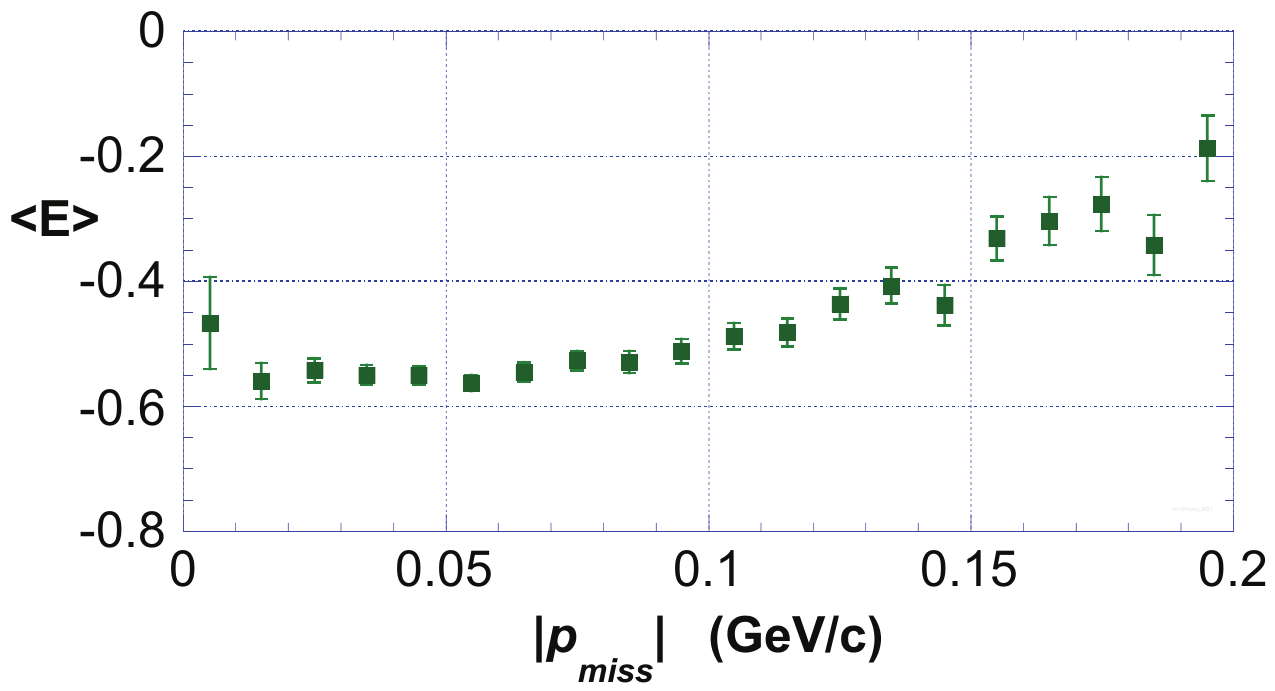}
\caption{The $E$ asymmetry for $\gamma+D\rightarrow \pi^- +p+(p_{miss})$, averaged over all angles and energies, as determined in the  BDT analysis. The uncertainties are statistical and are smallest near the peak of the $|p_{miss}|$ distribution ($\sim$0.06 GeV/c).}
\label{EvsPmiss} }   
\end{figure}

The final requirement common to all three analyses is the selection of events for which the neutron in deuterium is as close to {\it{free}} as possible, and the key parameter is the neutron momentum in the deuteron, or equivalently the reconstructed momentum of the undetected ({\it{spectator}}) proton, $P_{miss}$. Since different polarization observables may exhibit different sensitivities, we have chosen to determine the optimum threshold from the data itself. Studies with individual kinematic bins have shown a dilution of the $E$ asymmetry when the $|P_{miss}|$ threshold is increased above 0.1 GeV/c, but no statistically significant change for smaller values. When averaged over the full kinematic range, the mean value of $E$ is plotted in Fig.~\ref{EvsPmiss} as a function of missing momentum. This average again is stable below $0.1$ GeV/c but rises significantly at higher $|P_{miss}|$. Consequently, $|P_{miss}| \leq 0.1$ GeV/c has been required in all three analyses. (There is still a slight curvature below 0.1 in Fig.~\ref{EvsPmiss}, and polynomial extrapolations to $|P_{miss}|  =0$ would suggest a further 2\% correction. However, since at this level there could be some angle dependence, we have instead added this residual difference into the systematic uncertainties.)

The effect of the deuteron's $D$-state has been studied using an impulse approximation within the formulation of Ref. \cite{WSL15}, extended to include all relativistic transformations of the spin of the moving neutron \cite{Polyzou}. Dilution of the $E$ asymmetry can be significant whenever high spectator momenta are present, but is suppressed to negligible levels by the $|P_{miss}| \leq 0.1$ GeV/c requirement.

The combination of Monte Carlo simulations of the CLAS response to quasi-free $\gamma D\rightarrow\pi^- p (p)$, including Fermi motion, together with flux-scaled empty-cell data, reproduces the observed $|P_{miss}|$ distribution below 0.1~GeV/c, although deviations arise at higher momenta. Theoretically, the explicit effects of final state $NN$ interactions (FSI) and $\pi N$ rescattering on the $E$ asymmetry have been studied for the lower end of the g14 energy range \cite{Aren05,WSL15}, and found to be negligible for the $\pi^- pp$ final state, mainly because the dominant $I=1 ~pp$ wavefunction is orthogonal to the initial deuteron wavefunction. (In contrast, FSI effects are appreciable for $\pi^0 np$). 
From the above considerations, we regard the $E$ asymmetries reported here as reliable estimates for a {\it{free}} neutron.

\begin{figure}[b]
\center{
\includegraphics[scale=0.65]{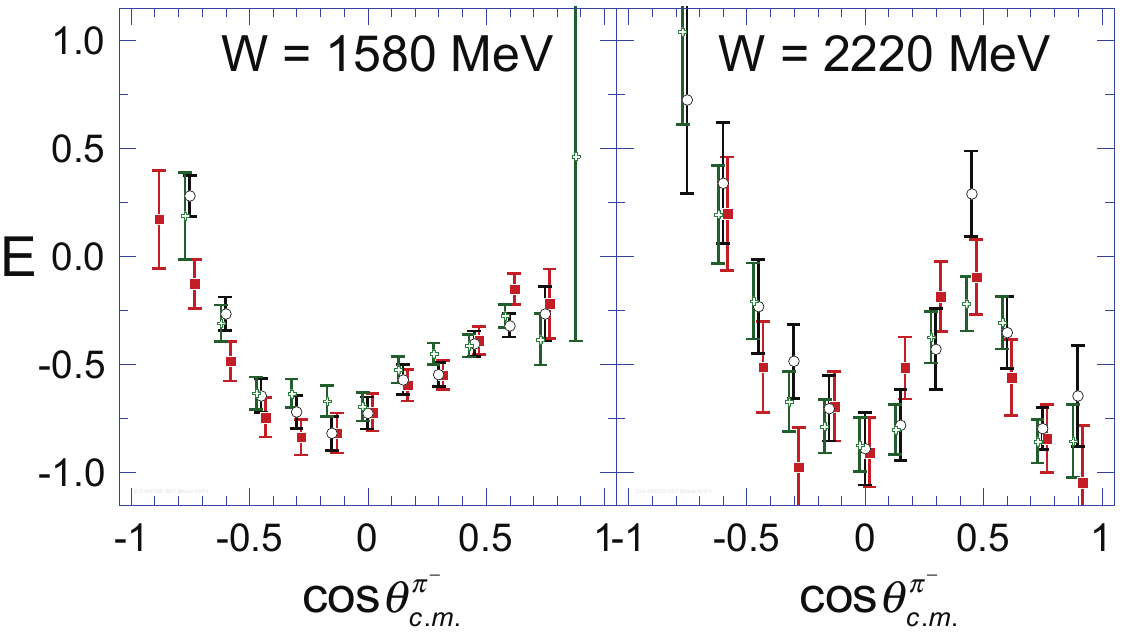}
\caption{Angular distributions of $E$ in the $\gamma n\rightarrow \pi^- p$ center of mass ($c.m.$) for two different $\pm 20$ MeV invariant mass bins, as determined from BKsub (black open circles), KinFit (red squares), and BDT (green crosses) analyses, respectively. KinFit and BDT points are shifted slightly in angle for clarity. }
\label{E_3methods} }   
\end{figure}

Asymmetries extracted from the BKsub, KinFit and BDT analyses are shown in Fig.~\ref{E_3methods} for two sample invariant mass ($W$) bins, near the low and high ends of the $W$ range. Results from the three data reduction methods are statistically consistent over the full energy range. 
A weighted average of the results from the three analyses has been used as the best estimate of the $\pi^- p$ {\it{E}} asymmetries. In calculating the net uncertainty, we have used standard methods to evaluate the correlations between the analyses \cite{schmel}, arising from the partial overlap of the sets of events retained by the three respective methods. 

\begin{figure*}
\center{
\includegraphics[scale=0.70]{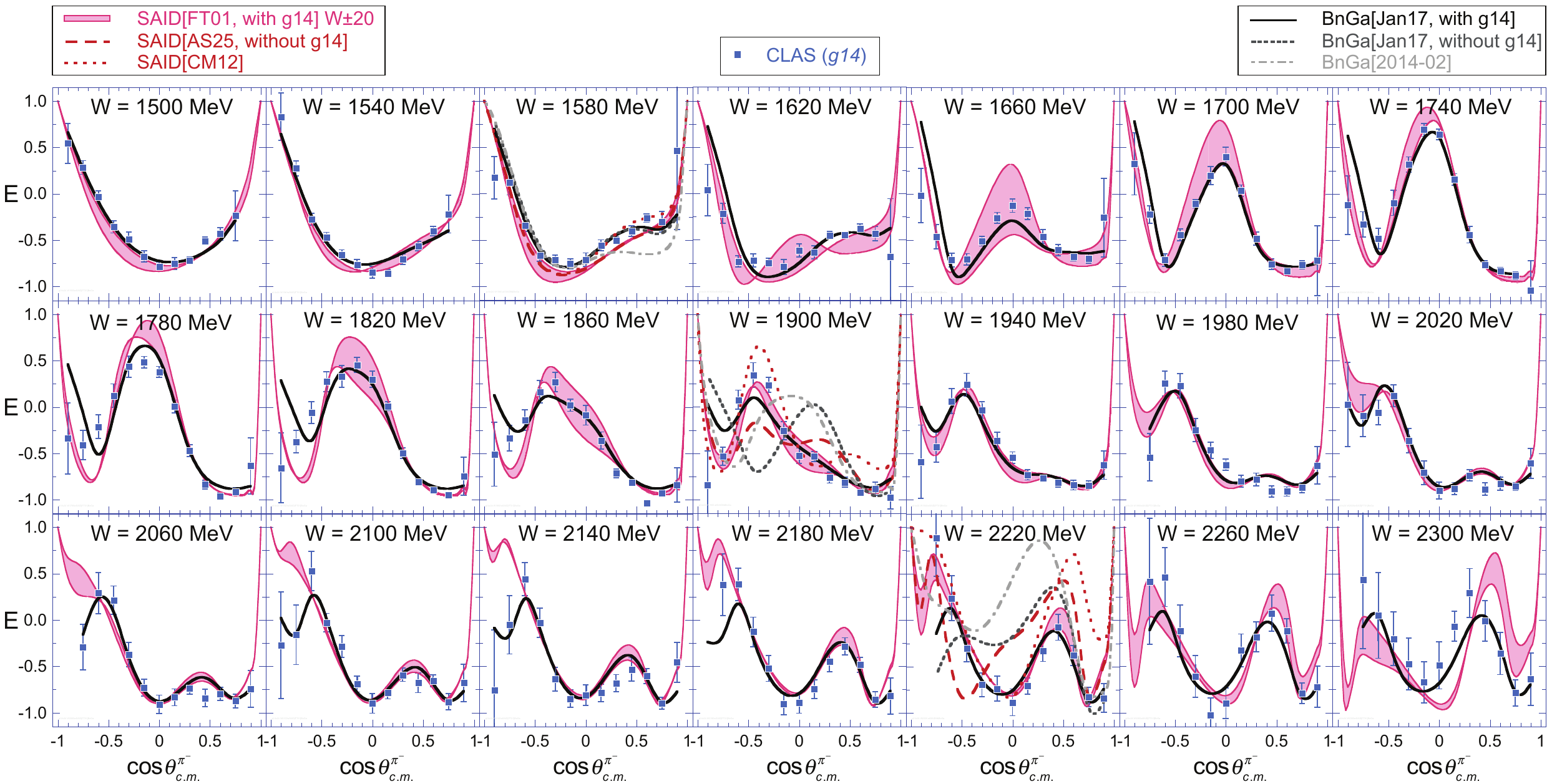}
\caption{$E$ asymmetries for $\vec{\gamma} \vec{n} \rightarrow \pi^- p$ (blue squares), grouped in $\pm 20$ MeV invariant mass ($W$) bins, shown with recent PWA fits that include these data: solid red curves from SAID \cite{SAID}, with shaded bands indicating variations across the energy bin; solid black lines from BnGa \cite{BnGa}. Also plotted at three $W$ values (1580, 1900, and 2220 MeV) are previous PWA solutions that did not include the present data set in the multipole search: red-dotted curves from SAID[CM12], based on all data up to 2012 \cite{CM12}; red-dashed curves from SAID[AS25], including all previously published data; grey dot-dashed curves from BnGa[2014-02], based on all data up to 2014 \cite{BnGa2014_02}; black short-dashed curves from a BnGa PWA using all previously published data.}
\label{E_g14wPWA} }   
\end{figure*}

Systematic uncertainties associated with event processing enter the three analyses in different ways, but total about $\pm$4\% (point to point) in each case. We assign an additional (point to point) uncertainty of $\pm$2\% to the uncorrected extrapolation to $|\vec{p}_{miss}| =0$. A relative uncertainty on polarization of $\pm$7\% (6.0\% target and 3.4\% beam) represents a scale uncertainty on the data set as a whole. The total systematic uncertainty is $\pm$8\%. 

Our final $E$ asymmetries are shown with statistical uncertainties in Fig.~\ref{E_g14wPWA}, grouped in $\pm 20$ MeV invariant mass bins. Numerical data files are available from Ref. \cite{CLAS_db}.

New Partial Wave Analyses (PWA) of $\pi$ photoproduction have been carried out, augmenting the neutron data base with these new $E$ asymmetries. New PWA from the George Washington University Data-Analysis group (SAID) \cite{SAID}, and new PWA from the Bonn-Gatchina (BnGa) group \cite{BnGa}, are shown as solid red and solid black curves in Fig.~\ref{E_g14wPWA}, respectively. Both provide very good representations of the new $E$ data. PWA combine results from many experiments at different energies, and this results in varying degrees of sensitivity to energy and angle. This is illustrated by the red bands whose width indicates the SAID variation across the energy bin.

The new $\pi^- p$ $E$ asymmetries have had a significant impact on multipole solutions. To illustrate their effect, we have plotted in Fig.~\ref{E_g14wPWA} the predictions from previous PWA solutions in a sample of three panels at low (1580~MeV), mid (1900~MeV), and high (2220~MeV) invariant masses. Predictions from the most recent on-line versions, SAID[CM12] \cite{CM12} and BnGa[2014-02] \cite{BnGa2014_02} are shown as the red-dotted and grey dash-dotted curves, respectively. Predictions from more recent PWA that include all currently published data \cite{newSAID_ref} (but exclude our $\pi^- p$ $E$ asymmetries) are shown as the red dashed and black short-dashed curves. While the earlier PWA solutions are close to the $E$ data at low energies, they become wildly disparate for $W$ above about $1800$ MeV.

\begin{figure}
\center{
\includegraphics[scale=0.8]{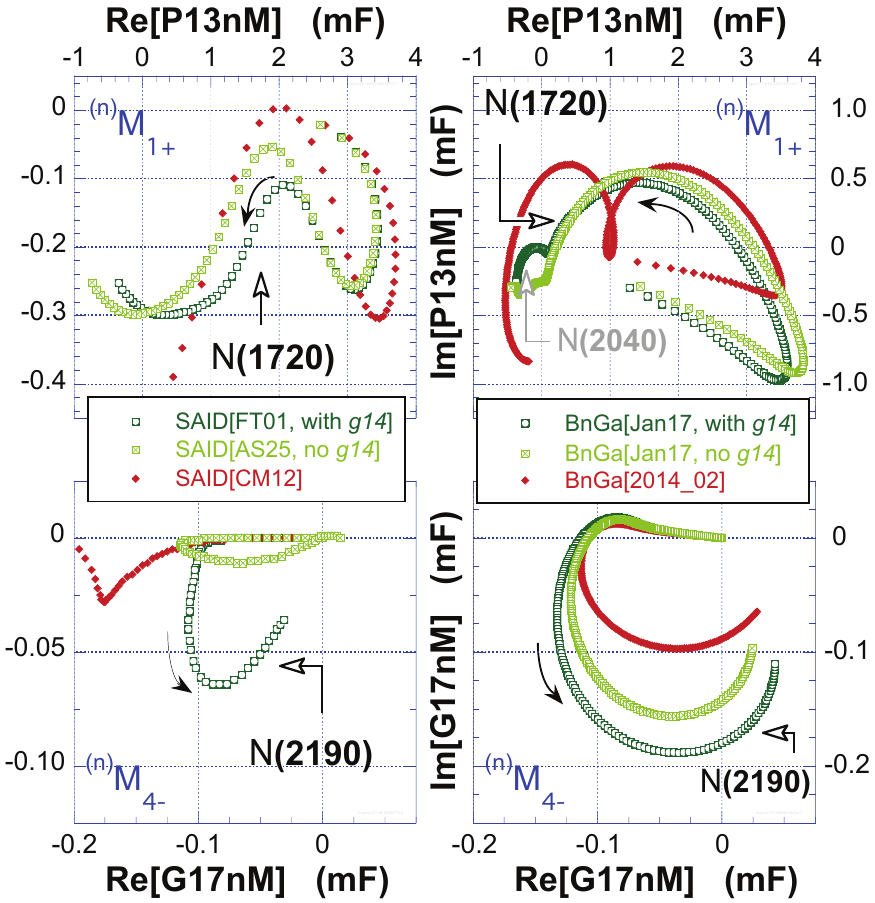}
\caption{Argand plots of the $P_{13}nM$ (top) and $G_{17}nM$ (bottom) multipoles from $\pi$-threshold to $W$=$2300$ MeV. Solid arrows indicate increasing $W$. SAID and BnGa PWA are shown in the left and right columns, respectively. As in the legend, red diamonds are on-line versions \cite{SAID,BnGa,CM12,BnGa2014_02}, light-green crossed squares are fitted to all previously published data, and dark-green circles augment these with the new $E$ asymmetries. }
\label{Argand} }   
\end{figure} 

As expected, the {\small $I=3/2$} partial waves, which can be determined entirely from proton target data, have remained essentially unaltered, while various {\small $I=1/2$} waves have changed substantially. As examples, in Fig.~\ref{Argand} we show Argand plots of the $(L^{\pi N})_{IJ}(n/p)E/M = P_{13}nM$ (top row) and $G_{17}nM$ (bottom row) partial waves. Both reveal the expected counter-clockwise phase motion near the $N(1720)3/2^+$ and $N(2190)7/2^-$ resonances, each ranked {\it{four star}} by the Particle Data Group \cite{PDG16}. Their corresponding centroids are indicated by open-black arrows. Recent PWA from  SAID and BnGa are plotted in the left and right columns, respectively. 

The $\gamma nN^*$ couplings can be expressed in terms of the transverse helicity amplitudes $A_n^h$, \cite{BnGa_Amp}. For the $N(2190)7/2^-$ resonance, the new $G_{17}$ BnGa multipoles (dark green squares in Fig.~\ref{Argand}) result in $A_n^{1/2} = +30 \pm 7$ and $A_n^{3/2} = -23 \pm 8$, in units of $10^{-3} $GeV$^{-1/2}$. This is significantly different from previous BnGa values of $-15 \pm 12$ and $-33 \pm 20$ \cite{PDG16, BnGa2013}, respectively. The corresponding new SAID PWA results in $A_n^{1/2} = -6 \pm 9$ and $A_n^{3/2} = -28 \pm 10$. The $G_{17}$ wave in previous SAID analyses had been too small to extract couplings. With the inclusion of the new $E$ asymmetry data, the SAID and BnGa $A_n^{3/2}$ amplitudes are in agreement.

From changes in the $P_{13}$ wave (top row of Fig.~\ref{Argand}), the SAID PWA has extracted new values of $A_n^{1/2} = -9 \pm2$ and $A_n^{3/2} = +19 \pm2$ for the $N(1720)3/2^+$. This is a significant revision from their previous values of $-21 \pm4$ and $-38 \pm7$, respectively \cite{SAID_SN11}. While changes are also evident in the BnGa PWA, the proximity of the $\rho$-threshold complicates this coupled-channel analysis, and revised couplings will be presented elsewhere. The new BnGa PWA also shows resonance-like phase motion near the mass of a {\it{one-star}}  $N(2040)3/2^+$ \cite{PDG16} (grey arrow in Fig.~\ref{Argand}). This state had not been explicitly included in their PWA and is now under study.

Several other {\small $I=1/2$} waves have also changed significantly. The influence of other data sets on these are currently under study, particularly since both charge channels are required to construct the isospin amplitude, {\small $\mathcal{A}_n^{I=1/2} =[\surd{2} \mathcal{A}_{\pi^- p} - \mathcal{A}_{\pi^0 n}]/3$}, and FSI are more problematic for the $\pi^0 n \tiny{~}p$ final state. New data on other observables are also expected in the near future, including an extensive set of cross sections from another CLAS experiment \cite{g13_dsg}, and further improvements in the determination of $N^*$ parameters can be anticipated.

In summary, the beam-target helicity asymmetry in the $\vec{\gamma} \vec{D} \rightarrow \pi^- p(p)$ reaction has been measured for the first time across the $N^*$ resonance region, and analysis constraints have been used to deduce the $E$ polarization asymmetry for an effective neutron target. Inclusion of these results in new PWA calculations has resulted in revised $\gamma nN^*$ couplings and, in the case of the $N(2190)7/2^-$, convergence among different PWA groups. Such couplings are sensitive to the dynamical process of $N^*$ excitation and provide important guides to nucleon structure models.

We would like to thank Dr. T.-S. H. Lee for many fruitful discussions and for his invaluable theoretical studies on the implications of analysis requirements. We are grateful for the outstanding assistance of the JLab Hall B and Accelerator technical staff. This work was supported by the US Department of Energy, Office of Nuclear Physics Division, under contract DE-AC05-06OR23177 under which Jefferson Science Associates operate Jefferson Laboratory, by the US National Science Foundation, by the Chilean Comisi\'on Nacional de Investigaci\'on Cient\'ifica y Tecnol\'ogica,  by the the French Centre National de la Recherche Scientifique and the the French Commissariat \`{a} l'Energie Atomique, by the Italian Istituto Nazionale di Fisica Nucleare, by the National Research Foundation of Korea, by the Scottish Universities Physics Alliance and by the United Kingdom's Science and Technology Facilities Council.

\end{document}